\documentclass[amsmath,amssymb,preprint,pre,aps]{revtex4-1}         

\usepackage{amssymb}
\usepackage{multirow}
\usepackage{amsfonts}
\usepackage{amsmath}
\usepackage{amssymb}
\usepackage{graphicx}
\usepackage{dcolumn}
\usepackage{times}
\usepackage{color}

\begin{document}

\title{Effective intermittency and cross-correlations in the Standard Map}

\author{G. Datseris}
\affiliation{Department of Physics, University of Athens, GR-15771 Athens, Greece}

\author{F. K. Diakonos}
\email[]{fdiakono@phys.uoa.gr}
\affiliation{Department of Physics, University of Athens, GR-15771 Athens, Greece}

\author{P. Schmelcher}
\email[]{pschmelc@physnet.uni-hamburg.de}
\affiliation{Zentrum f\"{u}r Optische Quantentechnologien, Universit\"{a}t Hamburg, Luruper Chaussee 149, 22761 Hamburg, Germany}
\affiliation{The Hamburg Centre for Ultrafast Imaging, Universit\"{a}t Hamburg, Luruper Chaussee 149, 22761 Hamburg, Germany}
\date{\today}

\begin{abstract}

We define auto- and cross-correlation functions capable to capture dynamical characteristics induced by local phase space structures in a general dynamical system. These correlation functions are calculated in the Standard Map for a range of values of the non-linearity parameter $k$. Using a model of non-interacting particles, each evolving according to the same Standard Map dynamics and located initially at specific phase space regions, we show that for $0.6 < k \leq 1.2$ long-range cross-correlations emerge. They occur as an ensemble property of particle trajectories by an appropriate choice of the phase space cells used in the statistical averaging. In this region of $k$-values the single particle phase space is either dominated by local chaos ($k \leq k_c$ with $k_c \approx 0.97$) or it is characterized by the transition from local to global chaos ($k_c < k \leq 1.2$). Introducing suitable symbolic dynamics we demonstrate that the emergence of long-range cross-correlations can be attributed to the existence of an effective intermittent dynamics in specific regions of the phase space. Our findings further support the recently found relation of intermittent dynamics with the occurrence of cross-correlations (F.K. Diakonos, A.K. Karlis, and P. Schmelcher, Europhys. Lett. {\bf 105}, 26004 (2014)) in simple one-dimensional intermittent maps, suggesting its validity for two-dimensional Hamiltonian systems.

\end{abstract}
\maketitle

\section{Introduction}

Dynamical systems with mixed phase space (PS) \cite{Chaos} are characterized by the coexistence of time scales which may differ by several orders of magnitude. The fast propagation in the chaotic sea is usually interrupted by long-time stickiness on local PS structures such as dynamical traps of hierarchically arranged islands, net-traps or dynamical traps of stochastic layers \cite{Zaslavsky2002}. Usually the dynamics in these regions is almost regular in contrast to the stochastic profile of the complete trajectory. Furthermore the PS geometry in the regions where regular and chaotic trajectories approach each other arbitrarily close is typically fractal and this leads to the presence of unusual statistical properties in the ensemble of the chaotic trajectories \cite{Karney1983}. This is mainly due to the fact that the dynamics is in this case pseudo-ergodic \cite{Zaslavsky1995} making impractical the use of trajectory based common tools, like for example the Lyapunov exponents or the correlation functions. In the context of Lyapunov exponents the main interest for a non-hyperbolic Hamiltonian system is to capture the large variation of the local instability due to the coexistence of chaotic and laminar phases along a reference chaotic trajectory \cite{Okushima2003}. A suitable tool for this purpose is provided by the finite-time Lyapunov exponents \cite{Dawson1994} which turn out to be much more sensitive to local PS structures than the global Lyapunov exponents and allow for a better understanding of the mixed PS dynamics. This is clearly demonstrated in \cite{Szezech2005} using as example Chirikov's Standard Map (SM) \cite{Chirikov1979} which is the prototype of a Hamiltonian system with mixed phase space. It provides an approximate description of several physical systems like the kicked rotor \cite{Lichtenberg1992}, the relativistic cyclotron \cite{Meiss1992} and the equilibrium configurations of the Frenkel-Kontorova model \cite{Aubry1983}. Being simple to simulate, two-dimensional and discrete in time, the SM has been extensively studied in the last decades with emphasis on diverse dynamical aspects such as anomalous diffusion and stickiness \cite{Rechester1980}, accelerating modes \cite{Ichikawa1987} and generating phase space partitions \cite{Christiansen1995}, which are closely related to the presence of local PS structures and their impact on mixed PS dynamics. 

For the case of correlation functions the implications of pseudo-ergodicity are more intricate, since these quantities contain an additional time scale (the delay time)  which implies a lower cut-off on the trajectory lengths used in the corresponding time-averaging. In fact for the correlation function to be representative for the PS dynamics it is required that the associated trajectory length is order(s) of magnitude larger than the maximum delay time. Thus the locality of the PS structures and the conditions of long time propagation are compelling factors making the use of correlation functions for exploring local PS dynamics a difficult task. On the other hand, in a class of Hamiltonian systems with PS dominated by stratification due to the presence of invariant KAM spanning curves, different PS regions may be dynamically disconnected. Thus, ergodicity may be partially restored within an isolated PS region and the definition of correlation functions within such a region is plausible. Usually, in this case, even after the destruction of the invariant spanning curves, remnants of the stratified structure survive in the form of a dense set of stability islands \cite{Robins2000}. Consequently, the stickiness on this net of dynamical traps induces large differences between the time scale characteristic for an ergodic covering of a single PS zone through a chaotic trajectory and the time scale needed to cross the destroyed spanning curve and enter in the accessible neighboring PS zone. One could naturally think that even in this case, despite the presence of global chaos, the definition of appropriate correlation functions sensitive to the local PS structures may be possible.

The aim of the present paper is twofold. Firstly we introduce a class of correlation functions possessing the ability to explore the dynamics in restricted PS regions for an arbitrary Hamiltonian system. Secondly, using these correlation functions we analyse the dynamics of the SM close to the transition from local to global chaos. In particular we focus on cross-correlation functions between trajectories with different initial conditions. Such a trajectory ensemble corresponds to a set of non-interacting particles each evolving according to the SM dynamics with the same $k$. We show that around the local to global chaos transition point, long-range cross-correlations develop, and subsequently, inspecting the trajectories in the ensemble used for the cross-correlation function calculations, we reveal the presence of effective intermittency in a symbolic representation of the associated dynamics. The observed intermittency is strong in the sense that the mean waiting time in the laminar region diverges. Thus, our analysis demonstrates that the previously recorded emergence of cross-correlations in a system of non-interacting particles following non-Hamiltonian 1-d intermittent dynamics \cite{Diakonos2014} is present also in a 2-D Hamiltonian system like the SM close to the transition from local to global chaos.

Our paper is organized as follows. In Section II we give the definitions of the correlation functions and we introduce the dynamical system we are considering in the subsequent analysis. Section III provides the numerical results of the correlation functions in the considered model for a dense set of values of the non-linearity parameter $k$ around the critical value $k_c$ signaling the transition from local to global chaos. In Section IV we analyse the obtained results by performing a detailed evaluation of the characteristic trajectories and introducing an appropriate symbolic dynamics to reveal the underlying effective intermittency. Finally Section V provides our concluding remarks. 

\section{Correlations in the Standard Map - model and basic observables}
The model we consider consists of $M$ non-interacting particles each evolving according to the dynamics of the Standard Map (\cite{Chirikov1979}):

\begin{equation}
\label{system}
\begin{split}
p^{(i)}_{n+1} &= p^{(i)}_n + k^{(i)} \sin \theta^{(i)}_n \\
\theta^{(i)}_{n+1} &=\theta^{(i)}_n + p^{(i)}_{n+1}
\end{split}
\end{equation}

\noindent
where $k^{(i)}$ is the control parameter of the non-linearity and $i=1,2,...,M$. We will here consider exclusively the case $k^{(1)}=k^{(2)}=.....k^{(M)}=k$ which simplifies the description significantly, reducing the study of the $M$-particle system to the study of an ensemble of initial conditions for a single particle. In this case
the particle index $(i)$ can be omitted in the description. Then the variables of the system, $\theta_n$ and $p_n$ stand for the single particle angular position and momentum at time instant $n$ respectively, both variables calculated with modulo 
$2\pi$. We restrict our analysis to the range $k \in [0,2]$. The system undergoes a transition from local to global chaos at the critical value $k_c=0.971635...$ \cite{Green1986}. At $k_c$ the last KAM curve with ratio $\phi$ (golden ratio) is destroyed and therefore chaotic orbits are not trapped between KAM tori any more, but they evolve across their remnants covering in principle all the available PS except the remaining islands of stability. Dynamically the trace of the destroyed regular PS structure is revealed in the stickiness of the chaotic orbits around these stability islands. As mentioned in the introduction it is likely to assume that the sticky evolution of the chaotic orbits in the immediate neighborhood of the remaining stability islands may introduce some correlations even between chaotic orbits originating from different initial conditions. To explore this scenario we will calculate the auto-correlation (AC) and cross-correlation (CC) functions of the considered system for the aforementioned $k$-values, focusing mainly on the $k$-region where the transition from local to global chaos takes place. 

For this task one usually employs the standard definition \cite{Chaos} for the correlation functions:

\begin{equation}
\label{cdef}
C_{x,ij}(m) = \lim_{N \rightarrow \infty} \left[ \frac{1}{N - m} \!\!\! \sum_{n=0}^{N-m-1} x_i(n)x_j(n+m) -\frac{1}{(N - m)^2} \!\!\! \sum_{n=0}^{N-m-1} \! x_i(n) \!\!\! \sum_{n=0}^{N-m-1} x_j(n+m) \right]
\end{equation}
where $N$ is the orbit's length, $x$ represents the PS variables 
$\theta$ or $p$, $m$ stands for the usual time-delay occurring in the correlation functions (CFs) and $i$, $j$ label the different sets of initial conditions 
$(\theta_0,p_0)$ that define the orbits $i$ and $j$. The diagonal entries $i=j$ in Eq.~(\ref{cdef}) determine the AC function which expresses the statistical similarity of a single trajectory at any two time instances differing by $m$. The CC function is obtained from the non-diagonal entries $i \neq j$ in Eq.~(\ref{cdef}) and expresses the respective statistical similarity between two different orbits (i.e. two orbits with different initial conditions) at any two time instances differing by $m$. The limit $N\rightarrow\infty$ is usually replaced by averaging over the entire PS requiring the system to be strongly ergodic, which is not for all the considered values of $k$ the case for the SM. In fact, in the most interesting region of $k$ $\in$ $[0.6,2.0]$, the PS is mixed. As a result the associated averaging contains qualitatively different dynamical components (chaotic, regular trajectories) and washes out properties (i.e. correlations) which characterize each dynamical component separately.  Thus, it is obvious that global averaging is not representative of the local structures of the phase space. 

Another property that disfavors the use of the global averaging is related to the symmetries of the SM. The PS of the Standard Map is point-symmetric around $(\pi,\pi)$ causing the fact that for each trajectory, denoted as $(\theta,p)$, there exists a partner trajectory $(\theta,p)_{ps} = (2\pi-\theta,2\pi-p)$ (where the index $ps$ is used to notice the point-symmetric partner of $(\theta,p)$). This implies that the contribution to the CC function of a pair of orbits $\{a,b\}$ and the contribution of the pair $\{a,b_{ps}\}$ with $b_{ps}$ the point-symmetric partner of $b$, will be exactly opposite canceling out in the averaging. Similarly, the contributions to the AC function of $a$ and $a_{ps}$ will also cancel out in the averaging having equal magnitude and opposite sign. Thus, as long as there coexist orbits in the ensemble with their point-symmetric counterparts, the averaging will always lead to vanishing auto- and cross-correlations.

In order to overcome both, the peculiarities of the averaging procedure originating from the fact that the ergodicity is not strongly satisfied, as well as the trivial behavior of the correlation functions due to the PS anti-symmetry, we introduce here a new class of correlation functions, the localized finite-time correlation functions (LFTCFs). Their aim is to extract information on the possible emergence of correlations due to the local structures in the phase space and they are defined as follows:

\begin{eqnarray}
\label{CLtrue}
LC^{(d)}_x(m) & = &\lim_{S_d \to \infty} \frac{1}{S_d}\sum_{i,j \in \mathfrak{C}^{(d)}}
C^{(d)}_{x,ij}(m) \nonumber \\
C^{(d)}_{x,ij}(m)& = & \frac{1}{N - m} \sum_{n=0}^{N-m-1}x_i^{(d)}(n)x_j^{(d)}(n+m)-\frac{1}{(N - m)^2}\sum_{n=0}^{N-m-1}x_i^{(d)}(n)\sum_{n=0}^{N-m-1}x_j^{(d)}(n+m) \nonumber \\
\text{where} & &\;\;\;\; \left\lbrace x_i^{(d)}(0), x_j^{(d)}(0) \right\rbrace \in \mathfrak{C}^{(d)} \subseteq PS~~~~~~~;~~~~~~x=\theta \; \text{or} \; p 
\end{eqnarray}

\begin{figure}[H]
\centering
\includegraphics[scale=0.55]{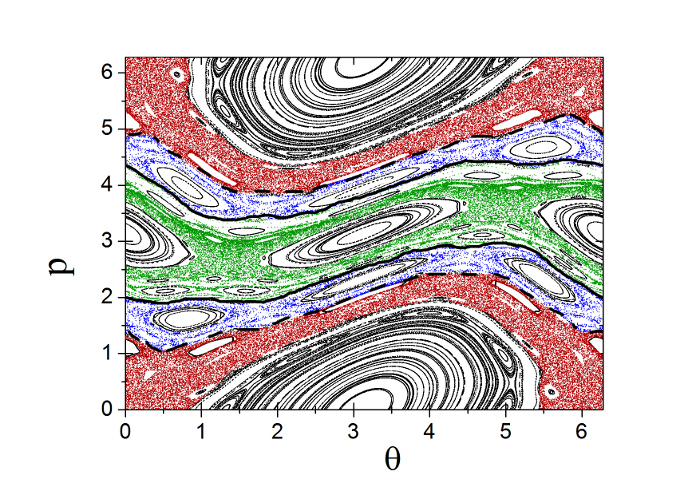}
\caption{(color online)~The mixed PS of the SM for $k=0.95$. The colored regions correspond to chaotic while the black regions to regular dynamics. The lines indicate the locations of $\Gamma$ (dashed) and $\Omega$ (solid) invariant spanning curves. The red colored region defines zone 1 in PS while the blue and the green colored regions define zones 2 and 3 respectively (see discussion in section III).} 
\label{fig:PS}
\end{figure}

The LFTCFs denoted by $LC$ in Eq.~(\ref{CLtrue}) are obtained by the usual definition of CFs (Eq.~(\ref{cdef})) by fixing the trajectory length $N$ in the ensemble to a finite value and averaging over different CFs, each calculated using trajectory pairs with initial conditions within a specific PS cell $\mathfrak{C}^{(d)}$. The index $d$ is used to indicate the PS domain where the cell $\mathfrak{C}^{(d)}$ belongs to (see below). In a similar manner as in Eq.~(\ref{cdef}) the case $i=j$ corresponds to the localized finite-time auto-correlation function (LFTACF) and $i \neq j$ corresponds to the localized finite-time cross-correlation function (LFTCCF). With $S_d$ we notice the total number of pairs of initial conditions used for the calculation of the CFs. In Eq.~(\ref{CLtrue})  $\mathfrak{C}^{(d)}$ should necessarily belong only to one of the two distinguishable domains covering the entire PS: 

\begin{itemize}
\item The \textit{chaotic domain} ($d=c$) which contains all the PS areas where local or global chaos is present. For $k \geq k_c$ the chaotic domain spans over the whole PS, excluding the stability islands.
\item The \textit{regular domain} ($d=r$) which contains all the areas of the PS that are either stability islands or spanning curves (periodic or quasi-periodic orbits).
\end{itemize}
As an example in Fig.~\ref{fig:PS} we show the PS of the SM for $k=0.95$. The chaotic domain corresponds to the colored regions while the regular domain to the black ones.
 
Since the definition of the LFTCFs requires the averaging over initial conditions in a specific PS-cell one can naturally ask about the dependence of the form of these CFs on the specific location or size of the cell within a domain as well as the length of the involved trajectories. This will be thoroughly discussed in section III. Obviously the LFTCFs allow us to analyze the correlations in each dynamical component (chaotic, regular) separately, taking also into account the impact of the local PS structure. However the cell choice requires some care since one has to ensure that it belongs entirely only to one of the two aforementioned domains. 

\section{Numerical Simulations}

For finite $S_d$ Eq.~\eqref{CLtrue} provides us with an estimate of the LFTCCF with initial conditions in the cell $\mathfrak{C}^{(d)}$. Of course the result will in general depend on both the finite time interval $N$ and the position of the cell $\mathfrak{C}^{(d)}$. This holds also for the LFTACFs. For a regular domain ($d=r$) the dominant frequencies define a characteristic time scale and $N$ can be chosen to be a multiple of the period corresponding to the largest frequency peak in the related power spectrum. In addition the regular motion does not generate diffusion and therefore the evolution of the initial cell $\mathfrak{C}^{(r)}$ remains localized in PS. Thus the validity of the terms "{\it finite time}" and "{\it local}" used in Eq.~\eqref{CLtrue} is straightforward for an ensemble of trajectories with initial conditions in a regular cell. 

In contrary, the validity of these terms is less clear when the cell of initial conditions lies within a chaotic domain ($d=c$). To illuminate this issue we explore in more detail the PS structure. For $0.55 <k < 0.85$ local chaos appears in the vicinity of the large regular island centered around $(\pi,0)$. For $k \in [0.85, 1.2]$ the PS is naturally divided into three zones which in practice are dynamically disconnected due to the presence of invariant KAM spanning curves or  their remnants. In Fig.~\ref{fig:PS} we show these zones colored red (zone 1), blue (zone 2) and green (zone 3) for $k=0.95$. Let $\Gamma$ be the last KAM spanning curve and its point-symmetric partner separating zones 1 and 2 (dashed lines in Fig.~\ref{fig:PS}) and $\Omega$ the last spanning curve and its point symmetric partner separating zones 2 and 3 (solid lines in Fig.~\ref{fig:PS}). $\Omega$ is the golden ratio KAM curve. At $k \approx 0.9164..$ the $\Gamma$ and at $k=k_c$ the $\Omega$ KAM curves are destroyed.
For $k\lesssim 0.91$ the term ``local" for the LFTCFs refers to dynamics within a single zone while the ``finite time" is irrelevant since an ensemble of chaotic trajectories with initial conditions within a zone remains there for very large time scales ($ \sim 10^8$ iterations).  Although the chaotic component of the PS becomes connected through the breaking of the spanning curves, there is always strong stickiness on the boundaries, and, as a result, the mean time needed for a trajectory starting within each one of these zones to enter in a neighboring zone, is very large ($ \sim 10^4-10^8$ iterations). In fact, even at $k=1$ where all spanning curves are broken, a trajectory starting in the red zone needs at least $10^3$ iterations to enter in the blue zone, while a trajectory starting in the blue zone needs at least $\sim 10^5$ iterations to enter into the green zone. To illustrate this in a more transparent way, we calculate the cumulative distribution function $F_T(t)= Prob(T \leq t)$ of the first passage time $T$ for a chaotic trajectory to cross up to time $t$, either the remnants of $\Gamma$ starting from zone 1, or the remnants of $\Omega$ starting from zone 2. In practice to find $F_T(t)$ we start with an ensemble of $10^5$ trajectories in a selected zone (1 or 2) and we determine the number of trajectories having first passage time $T$ smaller than $t$ normalizing by the total number of trajectories in the ensemble. 

\begin{figure}[H]
\centering
\includegraphics[scale=0.5]{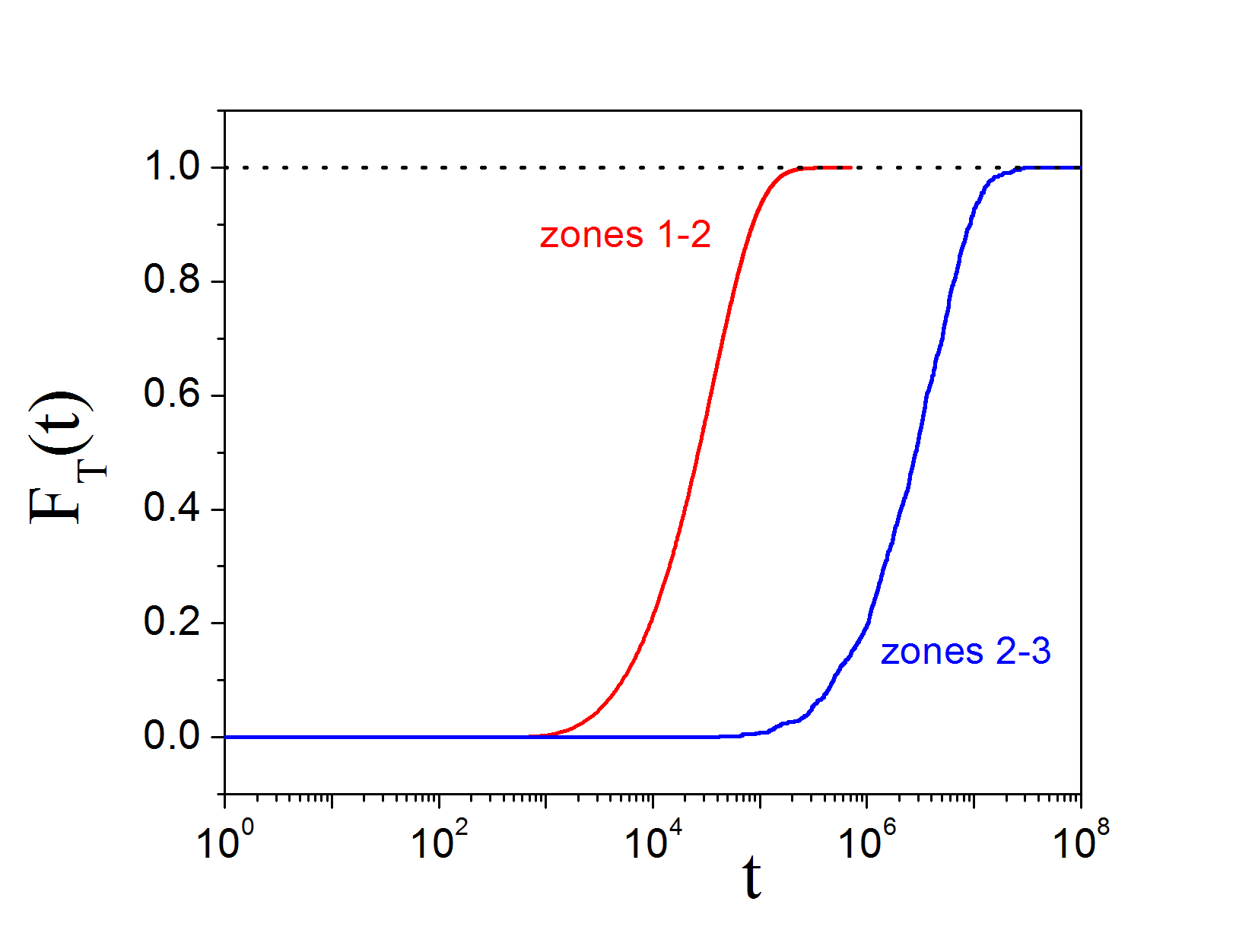}
\caption{(color online)~The cumulative distribution functions for the probability of a chaotic trajectory to cross up to time $t$ the second last spanning curve (red line) starting from a PS cell in the first zone (red region in Fig.~\ref{fig:PS}) and the last spanning curve (blue line) starting from a PS cell in the second zone (blue region in Fig.~\ref{fig:PS}). Both functions are calculated using $k=1$. The dotted line at $F_T(t)=1$ is plotted to guide the eye.} 
\label{fig:fvtdis}
\end{figure}

In Fig.~\ref{fig:fvtdis} we show $F_T(t)$ for the first passage time from zone 1 to zone 2 (red line) as well as the first passage time from zone 2 to zone 3 (blue line). As expected both curves saturate at $F_T=1$ for $t \to \infty$. We clearly observe the lower cut-offs at times $t_{12,min} \approx 10^3$ and $t_{23,min} \approx 10^5$ respectively. In addition the corresponding mean times (defined by $F_T(\langle t \rangle)=1/2$) are $\langle t_{12} \rangle=2.6 \cdot 10^4$ and $\langle t_{23} \rangle = 2.8 \cdot 10^6$. Thus, as dictated by the results shown in Fig.~\ref{fig:fvtdis}, there are well defined time scales which restrict the dynamics to a single zone. This justifies the use of the terms ``finite time" and ``local" in the correlation function of Eq.~(\ref{CLtrue}) even for chaotic trajectories with $k$ larger but close enough to $k_c$, so that the PS stratification into zones, as demonstrated in Fig.~\ref{fig:PS}, is still dynamically valid. For a large ensemble of trajectories with initial conditions in the red zone, choosing for example $N \approx 10^4$ in Eq.~(\ref{CLtrue}) it is guaranteed that the obtained correlation functions characterize mostly the local dynamics within this zone. A similar argument applies also for analogous ensembles of trajectories with initial conditions in the other two zones. The general conclusion is that LFTCFs are especially useful for the description of correlations at least in 2 cases: (i) when the PS of the dynamical system is stratified into zones by invariant spanning curves and within each zone local chaos is fully developed ($0.85 < k < 0.91$ for the SM) and (ii) close to the critical point for a dynamical system exhibiting a transition from local to global chaos ($0.91 < k < 1.2$ for the SM). Far beyond the critical point ($k \gg k_c$ for the SM), when the chaotic sea becomes almost homogeneous and dominating in PS while the remnants of the regular dynamics shrink significantly, the LFTCFs converge rapidly, i.e. for relatively small trajectory length $N$, to the usual correlation functions in Eq.~(\ref{cdef}) and become independent of the cell in the chaotic domain.  

Since the delay time $m$ should be much smaller than the length of the trajectories involved in the calculation of the LFTCFs we have to choose $N$ as large as possible in order to allow for a large variation of $m$. On the other hand, for $k \gtrsim k_c$, $N$ should be of the order of the minimum first passage time in the corresponding zone. For a given $N$, in order to ensure that the obtained numerical results are representative of the limit $S_d \to \infty$, we use as criterion the convergence of 2 significant digits when increasing $S_d$ by a factor of $10$. In addition the convergence with respect to the used length of the trajectories in the ensemble is tested by increasing $N$ by a factor of $2$. We found empirically that taking the values $S_d=5\cdot 10^5$ and $N=5\cdot 10^3$ in the calculation of \eqref{CLtrue} we achieve the required convergence. Thus, in the following discussion, if not stated differently, we present results using exclusively these parameter values for $N$ and $S_d$.

\subsection{Correlation functions in the regular domain}

\begin{figure}[H]
\centering
\includegraphics[scale=0.25]{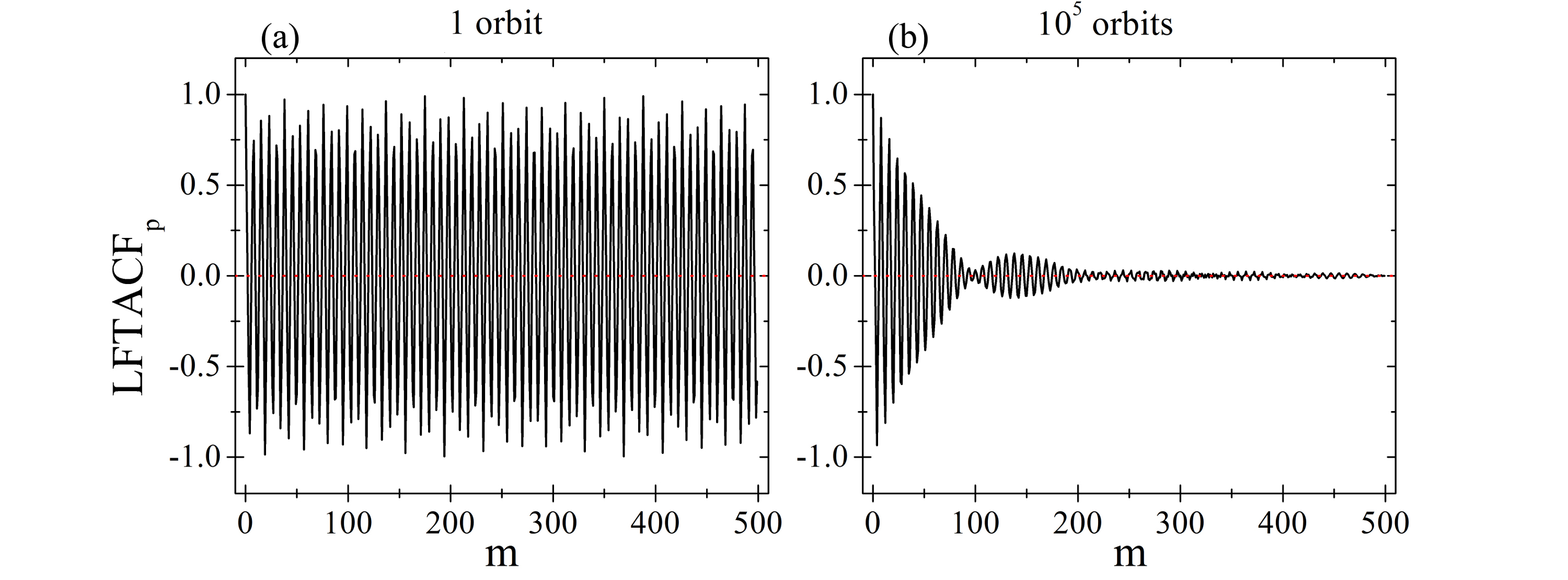}
\caption{Numerical demonstration of the cancellation of the correlations in the regular domain. Shown is the LFTACF for the $p$ component of orbits in a cell of size $0.5 \times 0.5$ within the stability island centered around $(\pi,0)$ when: (a) a single trajectory, and (b) $10^5$ trajectories are used in the averaging. The LFTCCFs (not shown) have similar form. Analogous results are obtained also for the LFTCFs of the $\theta$ variable.}
\label{fig:detun}
\end{figure}

First we discuss the results for LFTCFs using initial condition cells in the regular domain which turn out to possess some universal features. 
Cells within the regular domain contain quasi-periodic and periodic orbits. The corresponding LFTCFs perform oscillations around zero. Increasing $S_d$ the amplitude of the oscillations diminishes, tending to zero for large $m$. This is due to the fact that the ensemble averaging is over oscillating forms with slightly shifted frequencies which interfere destructively canceling each other. This effect is demonstrated in Fig.~\ref{fig:detun} where we plot the LFTACF of $p$ in (a)  
with $S_d=1$ and in (b) with $S_d=10^5$. The LFTCCF in the same cell (not shown here) converges also to an oscillatory form around zero, decaying much faster ($m \approx 30$) and having initially ($m < 30$) a very small amplitude ($O(10^{-3})$).

This is a typical behavior for all cells in the regular domain. Since the LFTCFs in the regular domain oscillate always around zero we will not consider them furthermore in the following.

\subsection{Auto-correlations in the chaotic domain}

As already mentioned, the LFTCFs may depend on the specific location of the cell within the chaotic domain. We start our analysis choosing as cell $\mathfrak{C}^{(c)}$ the square $[0,0.25]\times[0,0.25]$ in the $(\theta,p)$ plane which is in the immediate neighborhood of the 1$^{st}$ order unstable fixed point $(0,0)$ of the SM. We first calculate the LFTCFs in this cell and subsequently we explore their dependence on the location or the size of the cell.

\begin{figure}[H]
\centering
\includegraphics[scale=0.6]{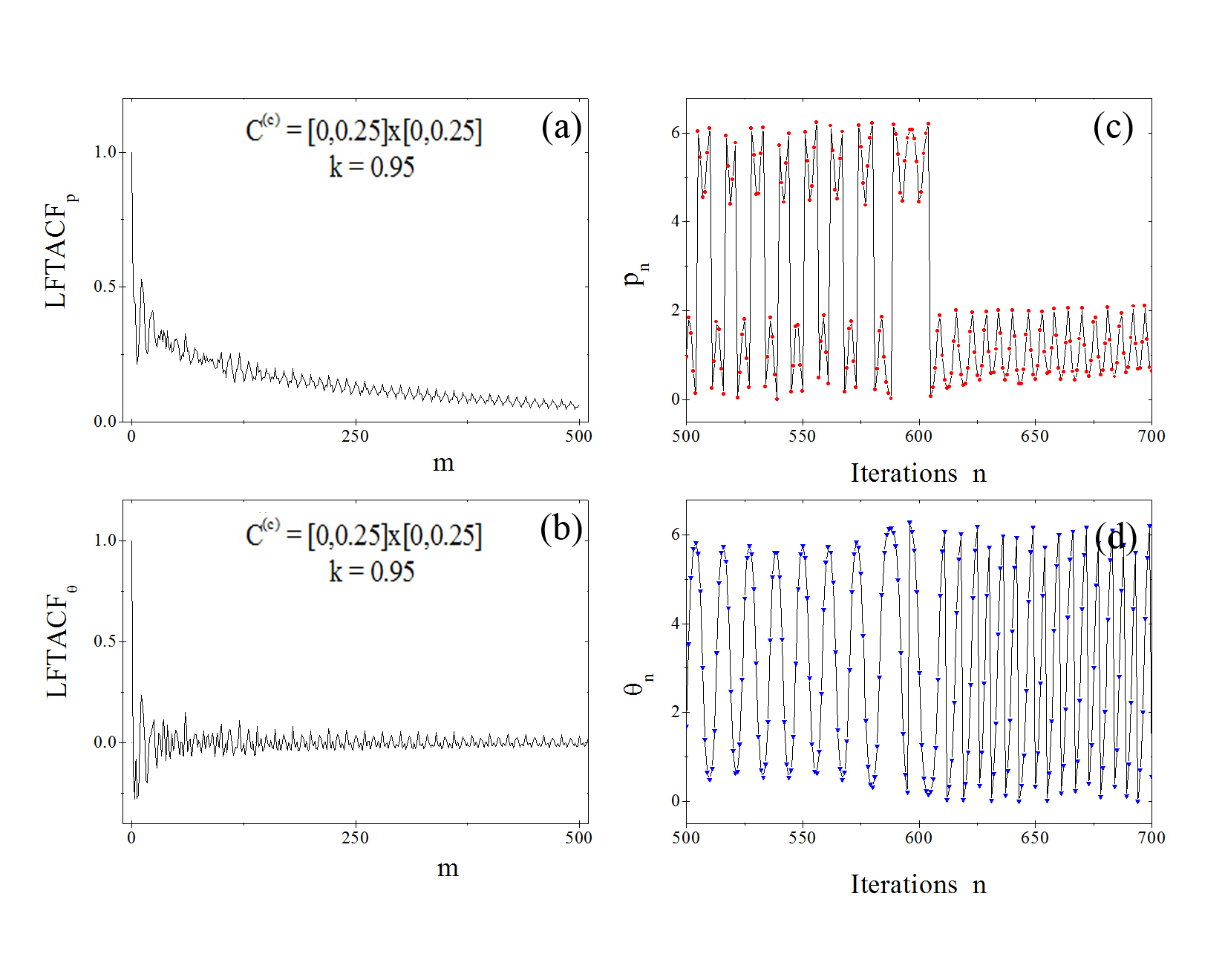}
\caption{(color online)~Left subfigures depict the average LFTACF for the cell $[0,0.25]\times[0,0.25]$ in $p$ ((a)) and $\theta$ ((b)) variables for $k=0.95$. Right subfigures are typical time-series for $p$ and $\theta$ with initial conditions in this cell. Subfigure (c) corresponds to $p_n$ and (d) to $\theta_n$ versus $n$.}
\label{fig:ACorbit}
\end{figure}

In figure \ref{fig:ACorbit}(ab) we present the LFTACF for $k=0.95$ using the phase space variables $p$ (a) and $\theta$ (b) respectively. Figs.~\ref{fig:ACorbit}(c,d) show component-wise ($p$ (c), $\theta$ (d)) a typical trajectory of the SM in zone 1 of the chaotic domain for this $k$-value.

Let us first discuss the behavior of the LFTACF$_{\theta}$, which is mostly small fluctuations around zero. This is easily explained by the form of the $\theta$-time-series. For the considered $k$-values $\theta$ performs an oscillatory motion. Consequently also the corresponding LFTACF$_\theta$ for a single trajectory possesses an oscillatory form. Averaging over different orbits will induce cancellations since the different oscillatory orbits vary in phase, amplitude and frequency. This is actually the case for all values of $k$, up to $k \approx 1.5$ where the orbits become strongly chaotic characterized by irregular fluctuations, independently of the cell location in PS. Following the same reasoning as it was the case for orbits in the regular domain, we conclude that the  LFTCCF$_\theta$ are fluctuations around zero and they also will not be considered furthermore in the following. Thus our subsequent analysis focuses only on the correlations of the $p$-component.

\begin{figure}[H]
\centering
\includegraphics[scale=0.55]{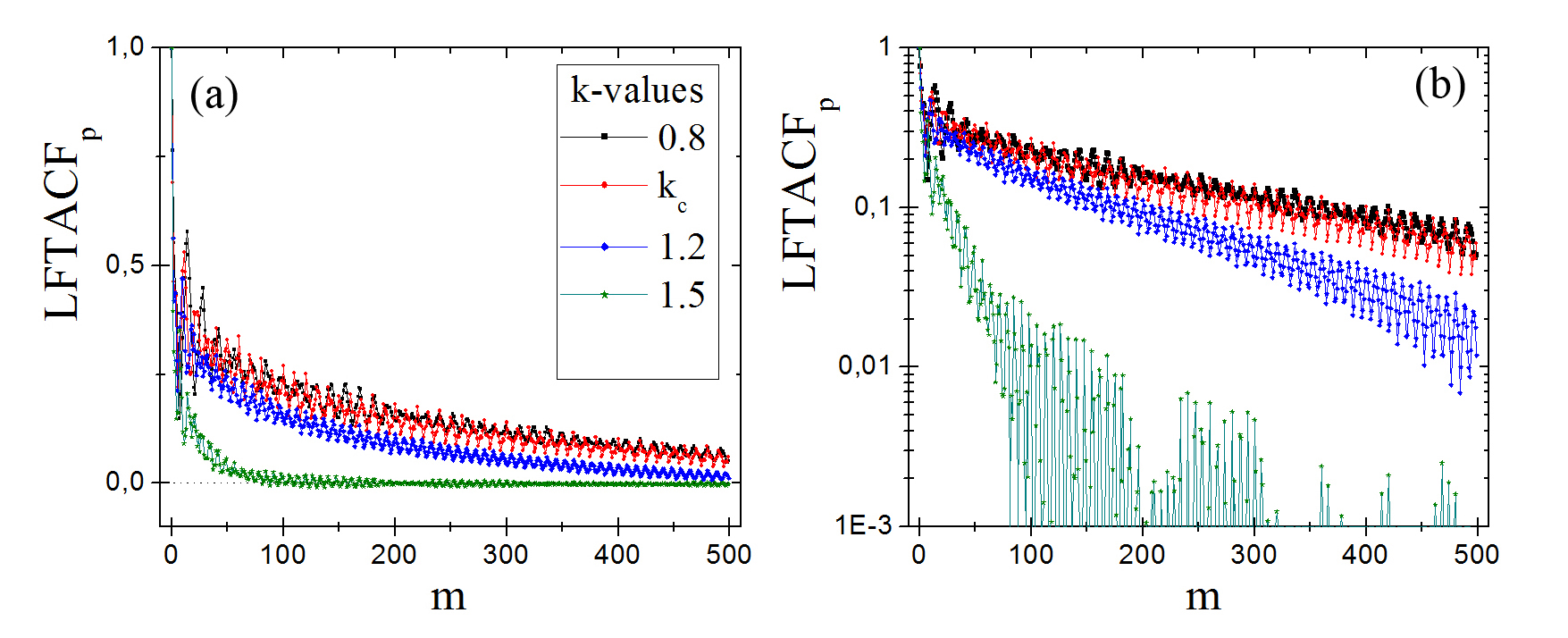}
\caption{(color online)~(a) LFTACFs for the $p$ variable in the cell $[0,0.25]\times[0,0.25]$ for a wide range of $k$-values plotted in linear scale. An ensemble of $S_d=5\cdot10^5$ orbits each of length $N=5\cdot 10^3$ is used. (b) The same plot in semi-log scale. The LFTACF$_p$ are normalized to one at $m=0$ by dividing the auto-correlation function of each trajectory in the ensemble with the corresponding standard deviation. }
\label{fig:ACk}
\end{figure}

\begin{figure}[H]
\centering
\includegraphics[scale=0.2]{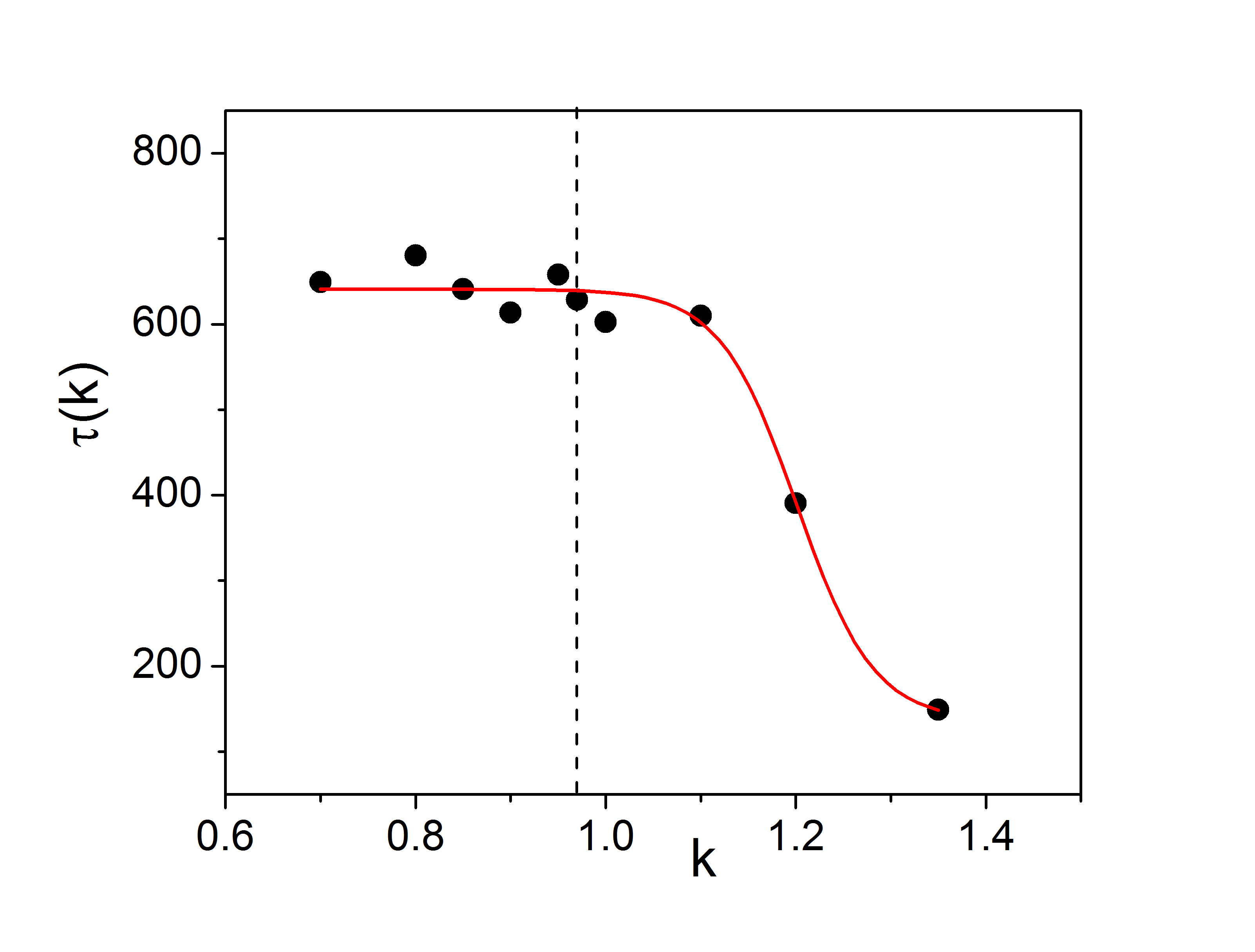}
\caption{(color online)~The characteristic exponents $\tau(k)$ of the LFTACFs (calculated using initial conditions in the PS cell $[0,0.25] \times [0,0.25]$) shown versus $k$ (black circles). The red solid line is an interpolating sigmoidal curve to guide the eye. The vertical dotted line indicates the location $k=k_c$.}
\label{fig:ACexponents}
\end{figure}

In contrast with the LFTACF$_{\theta}$, the auto-correlation of $p$, as observed in Fig.~\ref{fig:ACorbit}(a) is not trivial and the form of the corresponding trajectory (Fig.~\ref{fig:ACorbit}(c)) is qualitatively different. To explore further the auto-correlations of $p$ we calculated LFTACF$_p$ for several values of the parameter $k$. The results are summarized in Fig.~\ref{fig:ACk}. In all calculations we used initial conditions in the cell $[0,0.25] \times [0,0.25]$. 
In each case the convergence criteria are well satisfied. Considering the behavior in the semi-log scale shown in Fig.~\ref{fig:ACk}(b), we observe an exponential decay trend for all values of $k \in [0.6,1.5]$. Although the exponential behavior is evident, there are fluctuations of the LFTACFs around the exponential envelope. These fluctuations are not of statistical origin since they persist by increasing the number of trajectories in the ensemble by a factor $10^2$. In fact they can be attributed to the stickiness of the chaotic trajectories to the stable manifolds of the set of unstable periodic orbits surrounding the stability islands. The LFTACFs decay to zero significantly faster for all $k>1.2$ reaching an almost instantaneous decay for $k \gtrsim 1.5$. The reason for this behavior will be discussed  in depth in the next section where an orbit analysis is presented. Despite the oscillatory fluctuations of the LFTACFs it is useful to perform a fit using an exponential function
\begin{equation}
AC_p(m) = A_0 e^{-m/\tau(k)}
\end{equation}
to determine the characteristic exponent $\tau(k)$ for the exponential decay of the auto-correlations at different $k$-values. The resulting $\tau(k)$ is shown as a function of $k$ in Fig.~\ref{fig:ACexponents}.

For $k \in [0.6,k_c]$ where local chaos is present, the (inverse) exponents reside within a characteristic plateau, possessing an approximately constant value of $650$. Above $k_c$ where the transition from local to global chaos takes place, the exponents rapidly decrease, approaching zero for $k \approx 1.5$. In section IV we will see that the form of the chaotic trajectories changes significantly for $k > 1.5$ being characterized by irregular fluctuations, while this is not the case for $k < 1.5$. It will also become evident that the transition from local to global chaos induces this change in the orbits' structure, leading to vanishing LFTCFs for $k > 1.5$.

\begin{figure}[H]
\centering
\includegraphics[scale=0.2]{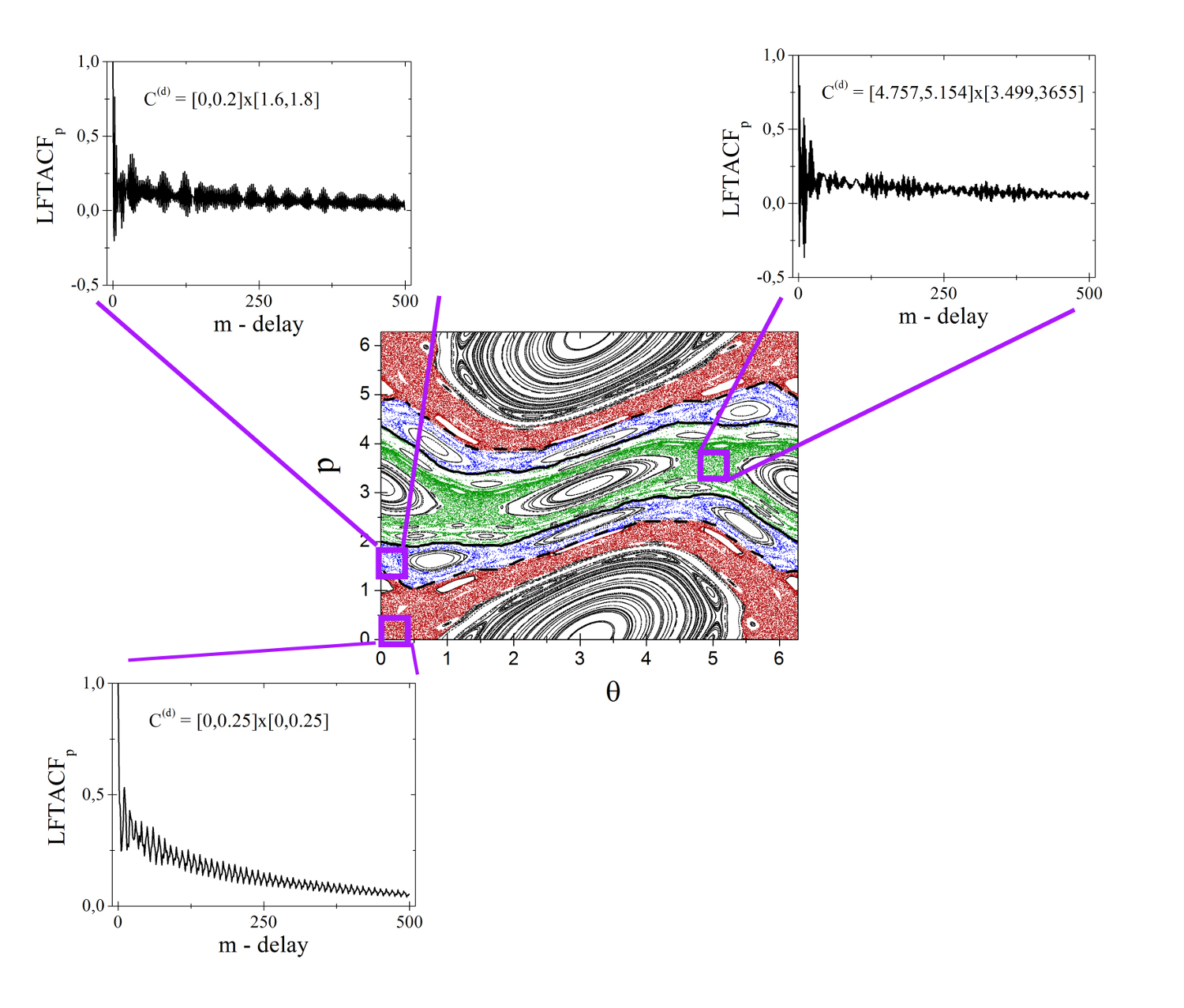}
\caption{(color online)~The LFTACFs belonging to the different zones of the chaotic domain at $k=0.95$.}
\label{fig:AClocation}
\end{figure}

Let us now turn to the dependence of the LFTACFs on the cell's location within the chaotic domain. This is demonstrated in Fig.~\ref{fig:AClocation}, showing LFTACFs calculated for different cell locations. 
It is evident that the LFTACFs depend on the zone where $\mathfrak{C}^{(c)}$ is located. Furthermore, we have checked the dependence of these results on the location of the cell $\mathfrak{C}^{(c)}$ within each zone. We have found that the form of the LFTACF does not change as we move $\mathfrak{C}^{(c)}$ within a zone, provided that we avoid the mixing of different PS domains (regular, chaotic).

\subsection{Cross-Correlations in the Chaotic Domain}

\begin{figure}[H]
\centering
\includegraphics[scale=0.335]{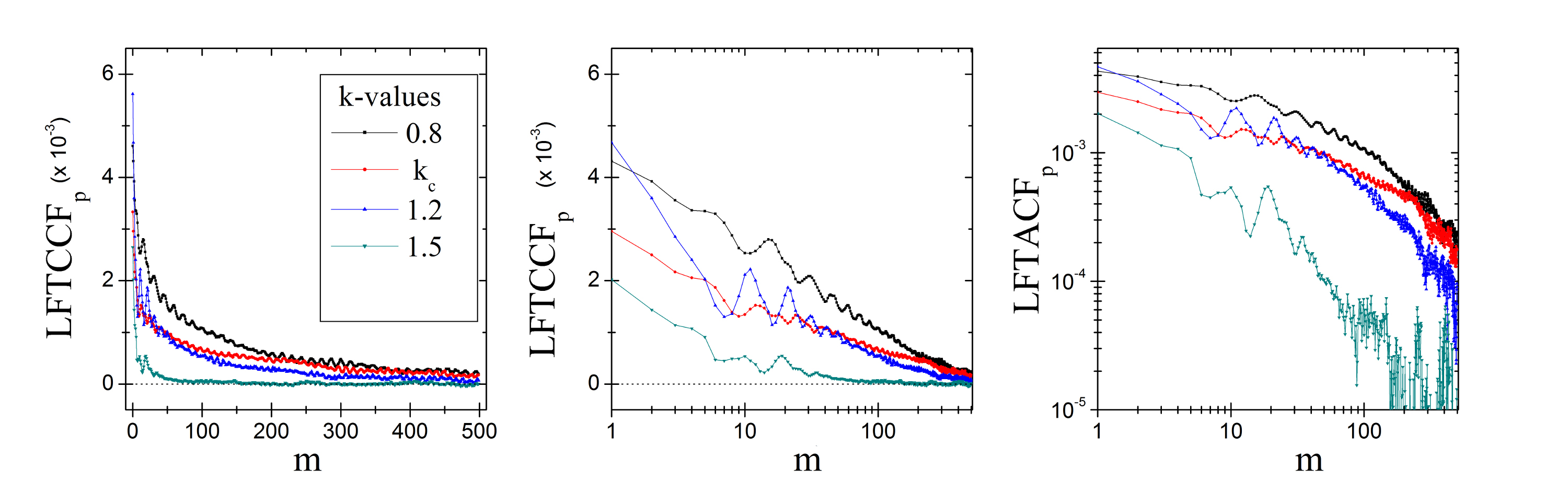}
\caption{(color online)~LFTCCFs of $p$ for an ensemble of $5\cdot 10^5$ chaotic trajectories with initial conditions in $\mathfrak{C}^{(c)}=[0,0.25]\times[0,0.25]$ (zone 1) obtained using $k=0.8$ (black line), $k=k_c$ (red line), $k=1.2$ (blue line) and $k=1.5$ (green line). (a) Linear-linear, (b) linear-log and (c) log-log scale. The cross-correlation function of each pair of trajectories in the ensemble is normalized by dividing with the root of the product of the standard deviations of those trajectories. } 
\label{fig:CCgeneral}
\end{figure}

Proceeding along the same lines as for the case of the auto-correlations, we analyze the LFTCCFs using as $\mathfrak{C}^{(c)}$ the cell $[0,0.25]\times[0,0.25]$. In Fig.~\ref{fig:CCgeneral} we present the LFTCCFs with initial conditions in this cell for various $k$-values.
Like in the LFTACFs case, the form of the LFTCCFs varies continuously with $k$ and for $k \gtrsim 1.5$ the LFTCCFs approach zero rapidly with increasing $m$  (see Fig.~\ref{fig:CCgeneral}(a)). Despite similarities, the structure of the LFTCCFs is quite different from that of the LFTACFs, as it becomes evident if we inspect the LFTCCFs in log-log (Fig.~\ref{fig:CCgeneral}(b)) and log-linear (Fig.~\ref{fig:CCgeneral}(c)) scales. Notice that the difference of approximately 3 orders of magnitude in the initial values of LFTCCF$_p$ and LFTACF$_p$ relies on the fact that the cross-correlation function of each pair of orbits in the corresponding ensemble, is not normalized to one at $m=0$ in contrast to the auto-correlation functions (see caption of Fig.~\ref{fig:CCgeneral}).

In Figs.~\ref{fig:CCgeneral}(b,c) it is clearly seen that, neglecting small amplitude fluctuations, the LFTCCFs show either a logarithmic or a power-law behavior, thus they possess long-range characteristics. In fact for $k<k_c$ the LFTCCFs can be better described as a logarithmic function of the delay $m$, while for $k_c<k<1.5$ they follow approximately a power-law. Since the long-range character of the LFTCCF weakens as $k$ increases beyond $k_c$, it is natural to ask the question whether it is related to the local structure of the chaotic domain induced by the last KAM spanning curves. Therefore in the following we will try to explain the emergence of long-range correlations by exploring in detail the SM chaotic dynamics close to $k_c$. In a recent study \cite{Diakonos2014} it has been shown that long-range cross-correlations can emerge in systems of non-interacting particles, each performing the same intermittent dynamics. Thus, one could ask if any kind of intermittent behavior within the chaotic domain of the Standard Map for these $k$-values is present (at least effectively) and generates these long-range cross-correlations. 

\begin{figure}[H]
\centering
\includegraphics[scale=0.45]{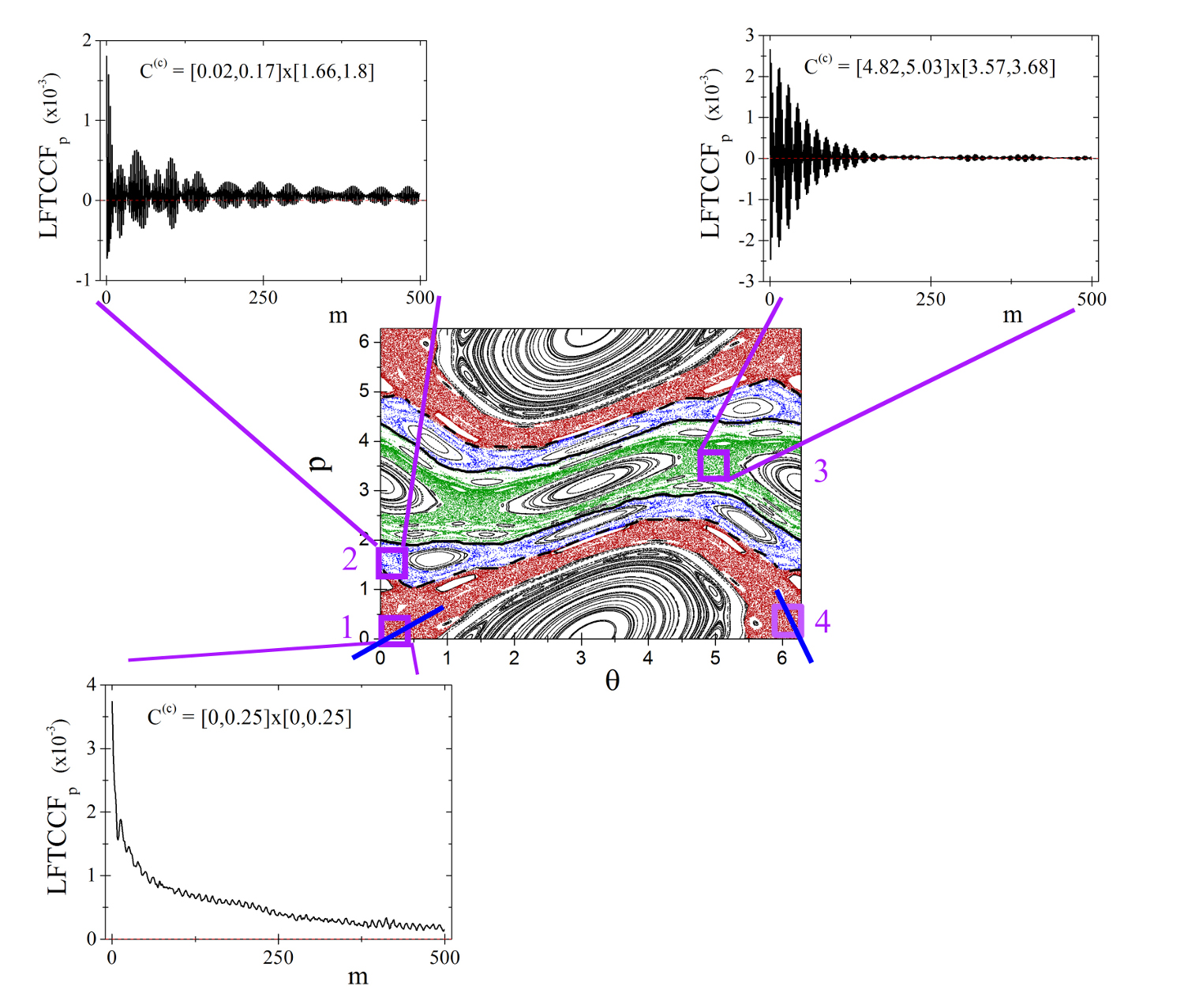}
\caption{(color online)~The LFTCCFs calculated for $k=0.95$ using PS cells located in different zones of the chaotic domain. For zone 1 we use $\mathfrak{C}^{(c)}=[0,0.25]\times [0,0.25]$, for zone 2 $\mathfrak{C}^{(c)}=[0.02,0.17]\times [1.66,1.8]$ and for zone 3 $\mathfrak{C}^{(c)}=[4.82,5.03]\times [3.57,3.68]$. Each cell is numbered with a violet-colored number. The blue lines above cells 1 and 4 correspond to the fixed point's $(0,0)$ eigenvectors (see end of section IV). The blue line above cell 1 represents the unstable eigenvector whereas the blue line above cell 4 represents the stable eigenvector.}
\label{fig:CClocation}
\end{figure}

Before we address the possibility of intermittency in the SM, a study of the dependence of the form of the LFTCCFs on the location of the cell $\mathfrak{C}^{(c)}$ is in order. Concentrating on the case $k =0.95$ we calculate as a first step the LFTCCFs using cells located in the different zones of the chaotic domain as presented by the different colored regions in Fig.~\ref{fig:PS}. The results of the corresponding simulations are summarized in Fig.~\ref{fig:CClocation}. 

In Fig.~\ref{fig:CClocation} it clearly can be seen that the LFTCCFs in zones 2 and 3 have a quite similar form while the LFTCCF in zone 1 behaves differently.  In the next section, this will be explained by exploring the dynamics in the neighborhood of the first order unstable fixed point $(0,0)$. As in the case of the LFTACFs, the results for LFTCCFs within each zone do not depend on the location of the cell $\mathfrak{C}^{(c)}$. However, we observe a very slow convergence when  
$\mathfrak{C}^{(c)}$ is located in the position ``4" within zone 1 (see Fig.~\ref{fig:CClocation}). This strange behavior will also be discussed at the end of the next section. Notice that in the case of zone 2, since the chaotic sea is very thin, one has to use particularly small cells for the calculation of the LFTCCFs. Due to the fact that the LFTCCFs in zones 2 and 3 are oscillating around zero while the LFTCCF in zone 1 has a long-range profile, in the next section we will focus exclusively on the emergence of cross-correlations in this zone.

\section{Intermittency and Orbit analysis}

Intermittency is characterized by long intervals of regular motion (laminar phase) of the trajectories, interrupted by chaotic bursts \cite{Manneville1979}. At a first glance there is no connection of the Eqs.~\eqref{system} describing the evolution in the SM with the normal form of intermittent dynamics \cite{Chaos}. Therefore it is not expected to find intermittent characteristics in the first iterate of the SM. On the other hand, an effective dynamics of the $p$-variable could generate such a behavior possibly in higher iterates. In order to inspect this hypothesis, a more in depth analysis of the trajectories in the parametric region $k \in [0.6,1.5]$, where the local to global chaos transition takes place, is required. Furthermore, as we have shown in the previous section, the long-range slowly decaying cross-correlations which could be associated with intermittent dynamics \cite{Diakonos2014} occur exclusively in SM trajectories evolving within zone 1 of the PS. Thus, in the following we will focus on dynamical characteristics of the chaotic trajectories in this PS region for $k \lesssim k_c$.

\begin{figure}[H]
\centering
\includegraphics[scale=0.92]{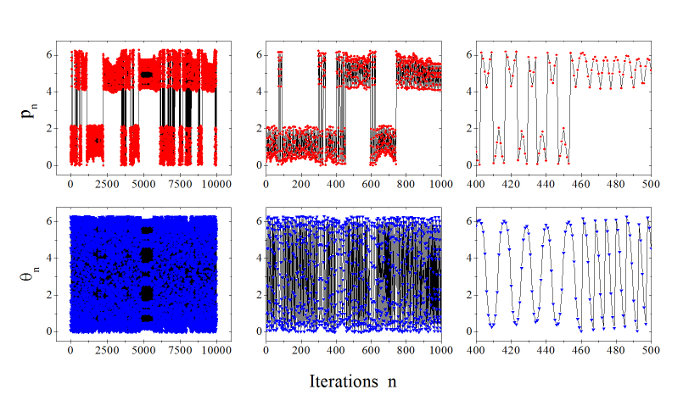}
\caption{(color online)~Time-series of the variables $p$ (a,b,c) and $\theta$ (d,e,f) of the Standard Map, for a typical orbit within the chaotic domain, for $k=0.95$ and initial conditions $(0.1,0.2)$. Shown are (a,d) the first $10^4$ iterations, (b,e) a zoom-in in the region $[0,10^3]$ and (c,f) a zoom-in in the region $[400,500]$ of this orbit.}
\label{fig:orbitanalysis}
\end{figure} 

\begin{figure}[H]
\centering
\includegraphics[scale=0.92]{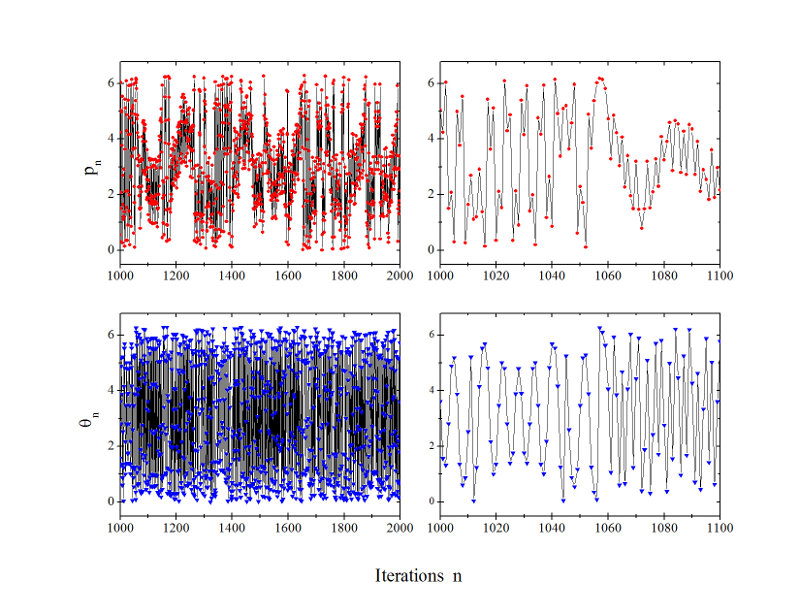}
\caption{(color online)~A typical trajectory for $k=1.8$ and initial conditions $(0.2,0.1)$. In the zoom-in it is clearly seen that the triangular substructure in the time-series of $p$ is not sustained and a laminar phase (as defined in section IV) does not exist.}
\label{fig:chaos}
\end{figure}

\begin{figure}[H]
\centering
\includegraphics[scale=0.8]{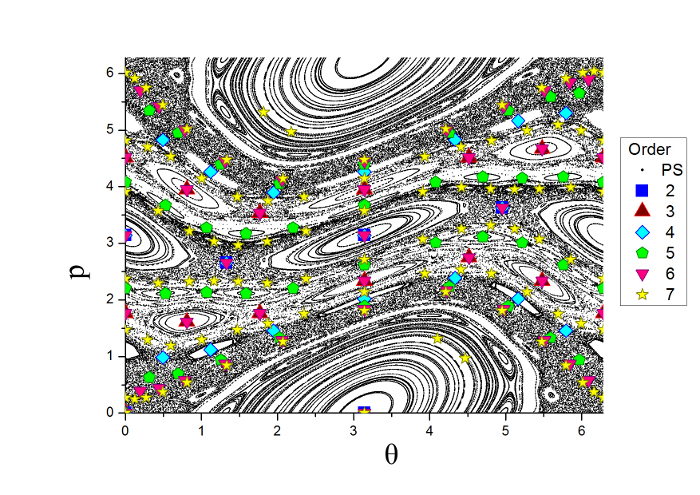}
\caption{(color online)~Fixed points of the Standard Map for $k=1.0$ found by the method introduced in \cite{Schmelcher1997}.}
\label{fig:fixedpoints}
\end{figure}

In this range of $k$-values the $p$-component of all trajectories in the chaotic region of zone 1 follows the behavior shown in Fig.~\ref{fig:orbitanalysis}(a-c). For $k$ beyond $1.5$ this structure is essentially destroyed  becoming gradually a completely irregular time-series of random fluctuations. A careful investigation of the orbit shown in Fig.~\ref{fig:orbitanalysis}(a-c), which is representative of all orbits in this chaotic domain for these $k$-values, leads to the following conclusions:

\begin{itemize}

\item The time-series of $p$ has a very intriguing structure. Considering as``down" the $p$-values below $\pi$ and as ``up" the $p$-values above it, we observe (see Figs.~\ref{fig:orbitanalysis}(a,b)) that the $p$ time-series consists of parts in which it is consecutively jumping up and down and parts in which it stays either only up or only down for a large number of iterations. This resembles roughly an intermittent time-series where the laminar phase corresponds to intervals for which the $p$-values are only up or only down while the chaotic bursts correspond to intervals with successive up-down (or vice versa) jumps of $p$.

\item Considering the $p$ time-series (Fig.~\ref{fig:orbitanalysis}(c)) at smaller time scales we observe that it consists of many sub-intervals each containing 4 to 7 points  (with $p$-values all above or all below $\pi$) forming an almost triangular structure. This property is typical for any $p$ time-series in zone 1 of the chaotic domain and it also does not depend on the time interval we zoom into. As demonstrated in Fig.~\ref{fig:fixedpoints} it originates from the unstable fixed points of order 4, 5, 6 and 7 of the SM. These are densely embedded in zone 1 and their manifolds affect the trajectories leading to the triangular structure formation. 

\item In fact, the generation of the triangular structures requires the synergy of the already mentioned unstable fixed points and a spanning curve (or its' remnants) acting as a separatrix, which keeps $p$ confined in zone 1. 
To illustrate this it is useful to consider the trace of the triangular structure in PS. Starting from the neighborhood of the PS point at $(0,0)$ the trajectory approaches the neighborhood of the PS point 
$(2 \pi, 0)$ in 4 to 7 iterations forming the triangular structure in the $p$ time-series. After reaching $(2\pi,0)$ the trajectory either returns to the neighborhood of $(0,0)$ in the next iteration, extending the duration of the laminar phase, or it jumps to the neighborhood of $(2 \pi, 2 \pi)$ signaling the beginning of a ``chaotic burst". The same process can also take place for trajectories involving the neighborhoods of the PS points $(2 \pi, 2\pi)$ and $(0, 2 \pi)$ which are the point-symmetric partners of $(0,0)$ and $(2 \pi, 0)$. During the chaotic burst interval of the $p$ time-series, the variable $\theta$ performs an oscillation from $0$ to $2 \pi$ and back (full cycle). When the $p$ time-series is in a laminar phase then the $\theta$ time-series consists of half cycles either from $2 \pi$ to $0$ (``up" laminar phase of $p$) or from $0$ to $2 \pi$ (``down" laminar phase of $p$). This behavior is clearly displayed in Fig.~\ref{fig:orbitanalysis}(f).

\item For $k>k_c$ the last spanning curve is destroyed. Therefore $p$ is not bounded any more and the triangular  structure is gradually degraded. However, for $k$-values up to $1.5$ the remnants of the separatrix still keep for long time intervals the $p$-trajectory trapped in zone 1 and therefore, although slightly deformed, the triangular structure is preserved. As a consequence the corresponding LFTCCFs can exhibit long range correlations for delay times comparable with the time scale needed for the $p$ trajectory to escape from zone 1. For example, at $k=1.2$ although the separatrix is destroyed, orbits with initial conditions in the neighborhood of $(0,0)$ and length about $N=10^4$ still exhibit the triangular structure without any significant change. However, orbits with $N=10^6$ have escaped through the traps created by the residues of the separatrix and contain large parts obeying irregular chaotic motion. This behavior is typical for all $p$-time series when $k > k_c$.

\item The structure of a typical PS trajectory changes completely for $k > 1.5$, as shown in Fig.~\ref{fig:chaos}. Zooming in we observe that the time-series is completely irregular in both $\theta$ and $p$ without any structure which could lead to a non-trivial behavior of the LFTCFs. This confirms the results presented in the previous section for $k$-values in this range.
\end{itemize}

The previous discussion suggests to represent the $p$ time-series by defining a suitable symbolic dynamics: every triangular unit consisting from 4 to 7 values of $p$ below $\pi$ is mapped to ``0" and every triangular unit consisting from 4 to 7 values of $p$ above $\pi$ is mapped to ``1". Obviously the opposite choice makes no difference. With this representation of the orbit we can easily define the laminar phase as a string of  consecutive 1s or 0s, i.e. 11111111.., and the chaotic bursts as strings that have alternating 1s and 0s, i.e. 10101010... Calculating the LFTCCFs with the symbolic time-series yields identical results as those in the previous section, scaled in time by a constant factor of $\approx 0.2$ since every 4-7 points of the original trajectory are mapped to one point in the symbolic time-series. 

\begin{figure}[H]
\centering
\includegraphics[scale=0.22]{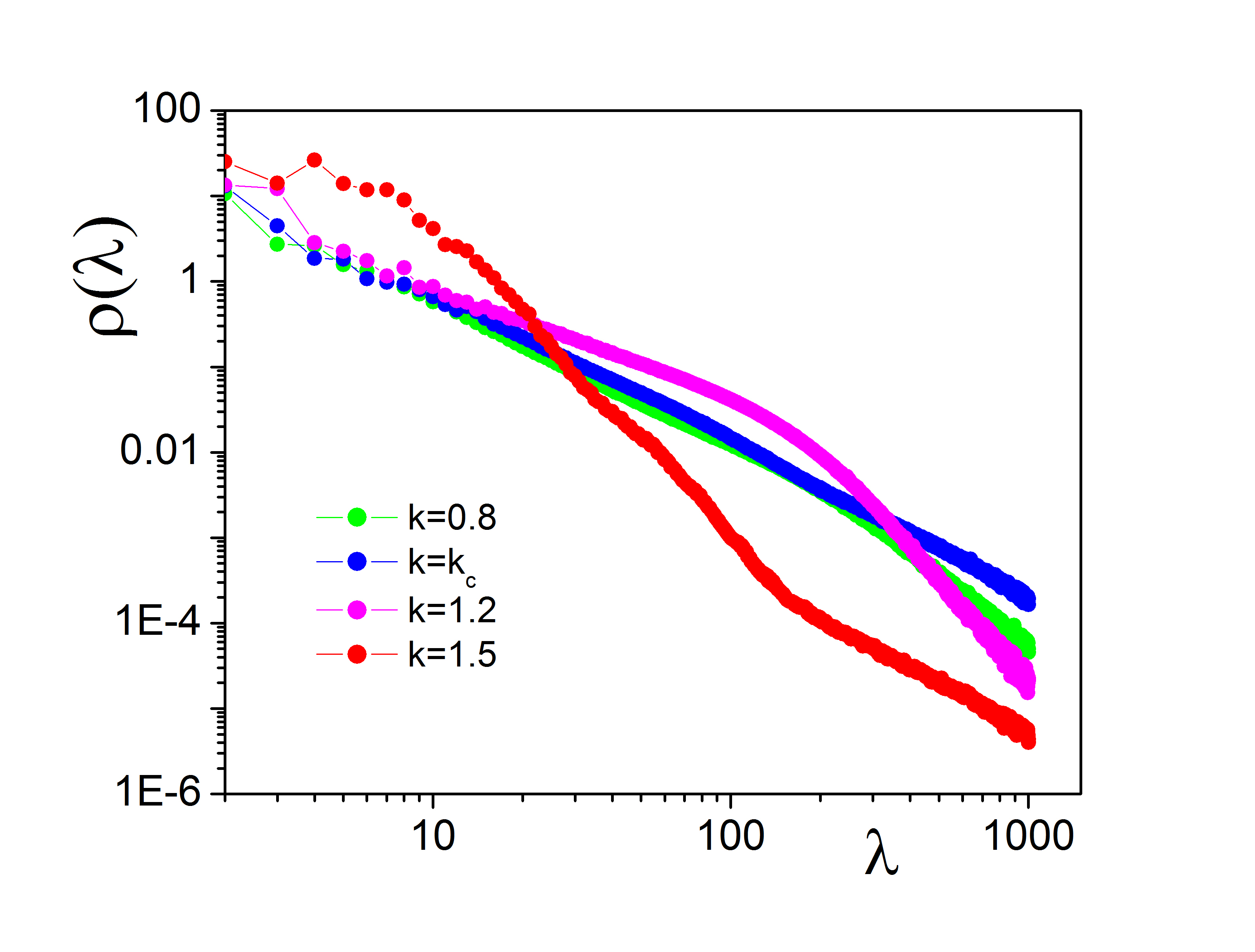}
\caption{(color online)~Laminar length distributions $\rho(\lambda)$ for various values of $k$, shown in log-log scale. We used $5\cdot 10^5$ trajectories with initial conditions in the cell $[0,0.25] \times [0,0.25]$ and verified the independence of the cell location in zone 1.}
\label{fig:laminar}
\end{figure}

It is now straightforward to calculate the lengths $\lambda$ of the laminar phases and to obtain the laminar length distributions. We used ensembles of $5\cdot 10^5$ trajectories of length $N=1.5 \cdot 10^4$ with initial conditions in the cell $[0,0.25] \times [0,0.25]$, for each $k$-value. The results are presented in Fig.~\ref{fig:laminar}. We checked convergence by increasing the number of trajectories in the ensemble by a factor of 10. The result coincides within 3 significant digits with that presented in Fig.~\ref{fig:laminar}. Remarkably enough it turns out that the laminar length distribution as defined above possesses some kind of universality. To verify this we recalculated this distribution by changing the size and the location of the cell of initial conditions in the chaotic domain of zone 1 and we found, for given $k$, the same result within a two significant digits accuracy.

The distributions shown in Fig.~\ref{fig:laminar} follow approximately power-laws for $k \in [0.6,k_c]$. Especially at $k=k_c$ the laminar length distribution is very close to a power-law. Beyond $k_c$ the distributions resemble a power-law but this form quickly dissolves, already having deviations at $k=1.2$. For $k > 1.5$ the laminar lengths as well as the associated symbolic dynamics lose their meaning, as the orbits no longer posses the aforementioned triangular structure.

The fact that for $k$ close to $k_c$ the laminar length distributions resemble a power-law is connected directly with intermittency. In \cite{Diakonos2014} it has been shown that the laminar length distributions for intermittent dynamics follow power-laws. There it was argued that the condition for the emergence of long range cross-correlations is that the mean laminar length $\left\langle\lambda\right\rangle$ diverges. By applying a power-law fit of the form
\begin{equation}
P(\lambda)=\frac{\beta}{\lambda^\alpha}
\end{equation}
to the laminar length distribution at $k=k_c$, we find $\alpha=1.85$. This implies that the average laminar length is 
\begin{equation}
\langle\lambda\rangle = \int_{2}^\infty P(\lambda)\lambda d\lambda = \int_{2}^\infty \lambda^{-0.85} d\lambda \rightarrow \infty
\label{symphony}
\end{equation}
which is in accordance with the statements in \cite{Diakonos2014}.

Two final comments are in order. As stated in subsection IIIC, in the calculation of the LFTCCF within zone 1 we observed a very slow convergence when the cell of initial conditions $\mathfrak{C}^{(c)}$ was chosen in the neighborhood of $(2 \pi,0)$ (cell with index ``4" in Fig.~\ref{fig:CClocation}). Here we will argue that this slow convergence is related to the presence of the stable manifold of the first order fixed point in this cell. To illustrate this it is useful to derive the corresponding stable and unstable eigenvectors in terms of the non-linearity parameter $k$: 
\begin{equation}
unstable: 
\left(
\begin{array}{c}
\frac{-k+\sqrt{(4+k)k}}{2} \\
1
\end{array}\right) \quad stable: \left(
\begin{array}{c}
-1 \\
\frac{-k+\sqrt{(4+k)k}}{2}
\end{array}\right)
\end{equation}
For $k\approx 1$ these eigenvectors are: 
\begin{equation}
unstable: 
\left(
\begin{array}{c}
0.62 \\
1
\end{array}\right) \quad stable:
\left(
\begin{array}{c}
-1 \\
0.62
\end{array}\right)
\end{equation}
Their direction is shown by the blue lines in Fig.~\ref{fig:CClocation}. According to this figure the cell with index ``1" contains the unstable manifold of $(0,0)$ while the cell ``4" contains the stable manifold. Although we were not able to determine the exact mechanism which induces the aforementioned slow convergence, we numerically proved that the stable manifold is the origin of this effect. To this end we calculated the LFTCCF for the time reversed dynamics (which transform the stable manifold to an unstable one) of the SM: 
\begin{equation}
\begin{split}
\theta_{n+1} &= \theta_n - p_n \\
p_{n+1} &= p_n -k\sin\theta_{n+1}
\end{split}
\label{timereverse}
\end{equation}
using initial conditions in the cell ``4". The obtained LFTCCF turned out to be identical to that using forward in time dynamics and initial conditions in the cell ``1". This clearly verifies our previous argumentation.

The last issue to be discussed concerns the understanding of the qualitatively different behavior of the LFTACFs and LFTCCFs for $k$-values around $k_c$. For this purpose let us consider more carefully the definition of the LFTCFs in Eq.~(\ref{CLtrue}). The sum $\sum_{i,j \in \mathfrak{C}^{(c)}}$ in Eq.~(\ref{CLtrue}) implies a summation over initial conditions $i$, $j$ either with $i=j$ (auto-correlations) or with $i \neq j$ (cross-correlations) where $x_i^{(d)}(0)$, $x_j^{(d)}(0)$ lie in the PS cell $\mathfrak{C}^{(c)}$. Focusing on the case $i \neq j$ and assuming that ergodicity is valid within a single PS zone, we could consider the chaotic trajectory $j$ as equivalent to the trajectory $i$ shifted forward in time by $n_{r,i}$ iterations, where $n_{r,i}$ is the recurrence time in $\mathfrak{C}^{(c)}$ for the trajectory $i$. Then we could approximately write $x^{(c)}_j(n+m) \approx x^{(c)}_i(n+n_{r,i} + m)$ and the sum over initial conditions occurring in the definition of the LFTCCFs could be written as:
\begin{equation}
\sum_{i,j \in \mathfrak{C}^{(d)}}=\sum_i \sum_{n_{r,i}} p(n_{r,i})
\label{sumdelay}
\end{equation}
where $p(n_{r,i})$ is the distribution of the recurrence times in $\mathfrak{C}^{(c)}$. Thus the averaging over different trajectory pairs is equivalent to an averaging over single trajectories like in the auto-correlation case, performing a random additional time shift following the recurrence time distribution and summing over all possible shifts. When the cell is located in the immediate neighborhood of the unstable fixed point $(0,0)$ the distribution $p(n_{r,i})$ can be approximated by the laminar length distribution attaining a power-law form.
If the cell $\mathfrak{C}^{(c)}$ is small, the distribution $p(n_{r,i})$ becomes to a good approximation independent of $i$ and one can use the notation $n_r$ for $n_{r,i}$. Then the LFTCF in Eq.~(\ref{CLtrue}) can be written as:
\begin{equation}
\begin{split}
LC^{(c)}_x(m)  =  \lim_{S_d \to \infty} \frac{1}{S_d} \sum_i \sum_{n_r} p(n_r) \left[
\frac{1}{N - m} \sum_{n=0}^{N-m-1}x_i^{(c)}(n)x_i^{(c)}(n+n_r+m)- \right. \\
\left.  \frac{1}{(N - m)^2}\sum_{n=0}^{N-m-1}x_i^{(c)}(n)\sum_{n=0}^{N-m-1}x_i^{(c)}(n+n_r+m)\right]
\end{split}
\label{ccapprox1}
\end{equation}
The term in the brackets is the auto-correlation function for the trajectory $x_i(n)$ evaluated at the delay time $n_r + m$. As we have shown in section IIIB the LFTACFs in zone 1 follow, for given $k$, an exponential form. The corresponding characteristic exponent $\tau(k)$ is defined as an ensemble property. Calculating the ACFs for individual trajectories forming the ensemble, we find that they can be approximated also by exponential functions of the delay $m$ however with an exponent $\tau_i(k)$ which depends on the initial conditions of the trajectory $i$. Thus, after performing the summation over $i$ we can approximate LFTCCF in zone 1 as:
\begin{equation}
LC^{(c)}_{x}(m) \approx \lim_{S_d \to \infty} \frac{1}{S_d} \sum_i\sum_{n_r} p(n_r) e^{-\frac{n_r +m}{\tau_i}}
\label{ccapprox2}
\end{equation}
where $p(n_r) \sim n_r^{-\alpha}$ with $\alpha \approx 1.85$ as dictated by the laminar length distribution.
The sum over $n_r$ can be performed leading to:
\begin{equation}
LC^{(c)}_{x}(m) \approx C_N \sum_i \tau_i^{1 - \alpha} e^{-\frac{m}{\tau_i}}
\label{ccapprox3}
\end{equation}
with $C_N$ a normalization factor. The sum in Eq.~(\ref{ccapprox3}), as a weighted infinite sum of exponentials, leads to the approximate power-law form of the  LFTCCF.  

\section{Concluding remarks}

We have introduced localized finite-time correlation functions as a suitable tool to explore the evolution of dynamical systems with mixed phase space. These correlation functions are sensitive to local phase space structures, quantifying the impact of stickiness due to phase space traps on the emergence of correlations in an ensemble of chaotic trajectories. Using as a prototype model the Standard Map we calculated these localized finite-time auto- and cross-correlation functions focusing on values of the non-linearity parameter $k$ in the regime of the local to global chaos transition. There the phase space is dynamically divided into three zones and the form of the aforementioned correlation functions differs in each zone. Specifically, in zone 1 (which includes the 1$^{st}$ order unstable fixed point) the auto-correlations show an exponentially decaying trend for a range of $k$-values around the critical $k_c \approx 0.971635..$, while the cross-correlations develop power-law tails signaling their long-range character. Around the transition point the dynamics of the Standard Map attains intermittent characteristics which can be revealed after a suitable symbolic dynamics is introduced. The pathway from this intermittent dynamics to the emergence of long-range cross-correlation is similar to the one introduced recently in \cite{Diakonos2014} for 1-D dissipative maps of Pomeau-Manneville type. Our results demonstrate that intermittency can effectively appear as a synergy of complicated phase space networks involving overlaps of unstable and stable manifolds of fixed points as well as partial confinement due to remnants of destroyed invariant spanning curves. Such a network may in general be established in a low dimensional dynamical system close to the critical point associated with the transition from local to global chaos. Furthermore it is confirmed that strong intermittency (even as effective dynamics) can generate long-range cross-correlations between chaotic trajectories as an ensemble property. This scenario can be easily transferred to the emergence of long-range cross-correlations between non-interacting particles as explained also in \cite{Diakonos2014} and may provide a useful manipulation tool for inducing collective behavior in non-interacting systems. 

\begin{acknowledgments}
We thank A. K. Karlis and B. Liebchen for fruitful and illuminating discussions. 
\end{acknowledgments}

\end{document}